\documentclass[traditabstract]{aa}
\usepackage{txfonts,natbib,graphicx,url}
\bibpunct{(}{)}{;}{a}{}{,}

\def\ltsima{$\; \buildrel < \over \sim \;$}
\def\simlt{\lower.5ex\hbox{\ltsima}}
\def\gtsima{$\; \buildrel > \over \sim \;$}
\def\simgt{\lower.5ex\hbox{\gtsima}}
\def\fullsrc{IGR J17498--2921}

\def\saxj{SAX J1808.4--3658}
\def\rxte{\it RXTE}

\begin{document}

\title{The discovery of the 401 Hz accreting millisecond pulsar {\fullsrc} in a 3.8 hr orbit}

\author{A.~Papitto\inst{\ref{inst1}} \and E.~Bozzo\inst{\ref{inst2}}
  \and C.~Ferrigno\inst{\ref{inst2}} \and T.~M.~Belloni\inst{\ref{inst3}}
  \and L.~Burderi\inst{\ref{inst1}} \and T.~Di Salvo\inst{\ref{inst4}}
  \and A.~Riggio\inst{\ref{inst5}} \and A.~D'A\`i\inst{\ref{inst4}}
  \and R.~Iaria\inst{\ref{inst4}}}

\institute{Dipartimento di Fisica, Universit\'a degli Studi di
  Cagliari, SP Monserrato-Sestu, KM 0.7, 09042 Monserrato,
  Italy\newline
  \texttt{e-mail:apapitto@oa-cagliari.inaf.it}\label{inst1} \and ISDC
  Science Data Center for Astrophysics of the University of Geneva,
  chemin d'\'Ecogia, 16, 1290, Versoix, Switzerland \label{inst2} \and
  INAF -- Osservatorio Astronomico di Brera, via E.~Bianchi 46, 23807,
  Merate, Italy \label{inst3} \and Dipartimento di Fisica,
  Universit\'a di Palermo, via Archirafi 36, 90123 Palermo,
  Italy \label{inst4} \and INAF -- Osservatorio Astronomico di
  Cagliari, Poggio dei Pini, Strada 54, 09012 Capoterra,
  Italy \label{inst5} }

\abstract{ 

We report on the detection of a 400.99018734(1) Hz coherent signal in
the {\it Rossi X-ray Timing Explorer} light curves of the recently
discovered X-ray transient, {\fullsrc}.  By analysing the frequency
modulation caused by the orbital motion observed between August 13 and
September 8, 2011, we derive an orbital solution for the binary system
with a period of 3.8432275(3) hr. The measured mass function, $f(M_2,
M_1, i)=0.00203807(8)$ M$_{\odot}$, allows to set a lower limit of
0.17 $M_{\odot}$ on the mass of the companion star, while an upper
limit of 0.48 $M_{\odot}$ is set by imposing that the companion star
does not overfill its Roche lobe. We observe a marginally significant
evolution of the signal frequency at an average rate of
$-(6.3\pm1.9)\times10^{-14}$ Hz s$^{-1}$. The low statistical
significance of this measurement and the possible presence of timing
noise hampers a firm detection of any evolution of the neutron star
spin. We also present an analysis of the spectral properties of
{\fullsrc} based on the observations performed by the {\it
  Swift}-X-ray Telescope and  the {\it RXTE}-Proportional Counter
Array between August 12 and September 22, 2011. During most of the
outburst, the spectra are modeled by a power-law with an index
$\Gamma\approx$1.7--2, while values of $\approx 3$ are observed as the
source fades into quiescence.

}

\keywords{ stars: neutron --- stars: rotation --- pulsars: individual ({\fullsrc}) --- X-rays: binaries }

\titlerunning{Discovery of a 401 Hz accreting ms pulsar in a 3.8 hr orbit}
\authorrunning{A.~Papitto et al.}
\maketitle

\section{Introduction}

 The discovery of a 401 Hz coherent signal from the X-ray transient
 {\saxj} \citep{WijvdK98} confirmed that old neutron stars (NS) in low
 mass X-ray binaries (LMXB) are rapidly rotating objects, spun-up by
 the accretion of angular momentum from matter in-falling through a
 disc. Sources showing a coherent signal with a period of few ms are
 called accreting millisecond pulsars (AMSP); they are all relatively
 faint X-ray transients attaining peak X-ray luminosities of a few
 $\times\:10^{36}$ erg s$^{-1}$, with outbursts lasting up to a few
 months.

Here we report on the discovery of 401 Hz pulsations from the X-ray
transient {\fullsrc}, which was first detected on 2011 Aug 11.9 thanks to
an \textsl{INTEGRAL} observation \citep[][the epochs in this paper
  are referred to the barycentric dynamical time, TDB]{Gbd11}. A
subsequent {\it Rossi X-ray Timing Explorer} ({\rxte}) observation
allowed the discovery of a coherent signal at 401 Hz \citep{Ppt11a},
bringing to 12 the number of known AMSPs\footnote{The properties of
  the intermittent pulsars, Aql X-1 \citep{casella08} and SAX
  J1748.9--2021 \citep{altamirano08}, seem distinct enough not to
  count them in this number.}.  The source position was determined by
follow-up \textsl{Swift} \citep{Bzz11a} and \textsl{Chandra}
\citep{Chk11} observations.  The latter estimate,
RA=17$^h\:49^m\:55^s$.35, DEC=$-29^{\circ}\:19'\:19.6"$, with an
uncertainty of 0.6$''$ at a 90\% confidence level, is considered
here. Bursts were detected by \citet{Frr11} during \textsl{INTEGRAL}
observations, while \citet{Lnr11} found oscillations at a frequency
consistent with the spin frequency during a burst observed by
\textsl{RXTE}, as well as a photospheric radius expansion episode,
from which a distance estimate  of 7.6 kpc was derived.
The transient returned to quiescence on 2011
  Sep 19, after a 37 day-long outburst \citep{linares11}.

The analysis of the \textsl{RXTE} observations performed during the
first few days of the outburst allowed \citet{Mrk11} to propose a set
of candidate orbital solutions, and \citet{Ppt11b} to give the first
preliminary solution of a 3.84 hr orbit of the NS. Here we
present the first analysis of the {\fullsrc} observations performed by
\textsl{RXTE} and \textsl{Swift}, focusing on the
properties of the 401 Hz signal   observed by
\textsl{RXTE} and deriving a full orbital and timing solution for
the pulsar.

\section{Observations and data analysis}
\label{sec:obs}

We processed data obtained by the \textsl{Swift}/X-ray Telescope
\citep[XRT; 0.2--10 keV,][]{gehrels04} between 2011 Aug 12.6 and
Sep 22.7 (MJD 55785.6--55826.7), for an exposure of 48.3
ks,  using standard procedures \citep{burrows05} and the
latest calibration files available at the time of the analysis
(September 2011). We considered observations performed in photon
counting mode, retaining the full spectroscopic and imaging
capabilities of the XRT with a time resolution of 2.5 s, and windowed
timing mode, which preserves only one imaging axis to allow a read out
every 1.7 ms. According to the recommendations of the XRT calibration
team\footnote{see
  http://heasarc.gsfc.nasa.gov/docs/heasarc/caldb/swift/docs/xrt/\\SWIFT-XRT-CALDB-09\_v16.pdf},
we produced 1--10 keV energy spectra grouping channels to contain at
least 50 counts each, adding an estimated systematic error of 3\% to each bin.

 Reduction and analysis of the data obtained by the
 \textsl{RXTE}/Proportional Counter Array \citep[PCA; 2--60
   keV;][]{jahoda06} between 2011 Aug 13.1 to Sep 22.4
 (MJD 55786.1--55826.4), for an exposure of 94.2 ks, was carried
 out with the tools available in HEASOFT (ver. 6.11). Spectra were
 extracted from data encoded in Standard 2 mode (time resolution of 16
 s and 128 spectral channels).  We retained only photons collected by
 the top layer of the Proportional Counter Unit (PCU) 2 in the 2.5--30
 keV band, adding a systematic error of 0.5\% to each spectral
 channel, to get the most reliable description of the PCA energy
 redistribution\footnote{see
   http://www.universe.nasa.gov/xrays/programs/rxte/pca/doc/rmf/}. The
 background was estimated with the 'bright' model. A timing analysis
 was performed on 2--60 keV photons observed by all the PCUs switched
 on during each of the observations, to achieve the highest possible
 quality counting statistics. Data processed in fast encoding modes
 such as event, single bit (both with 122 $\mu$s time resolution), and
 good xenon (1 $\mu$s time resolution) are considered.

Since in these paper we deal with the source properties during its
\textsl{persistent} (i.e., non-bursting) emission, we discard 5s of
data before each of the seven bursts detected by \textsl{RXTE}, as
well as before the two observed by \textsl{Swift}, and an interval of
variable length depending on the burst properties ($\sim60$ s), after.

\subsection{Light curves and spectra}
\label{sec:spectra}

\begin{figure}
\resizebox{\hsize}{!}{\includegraphics{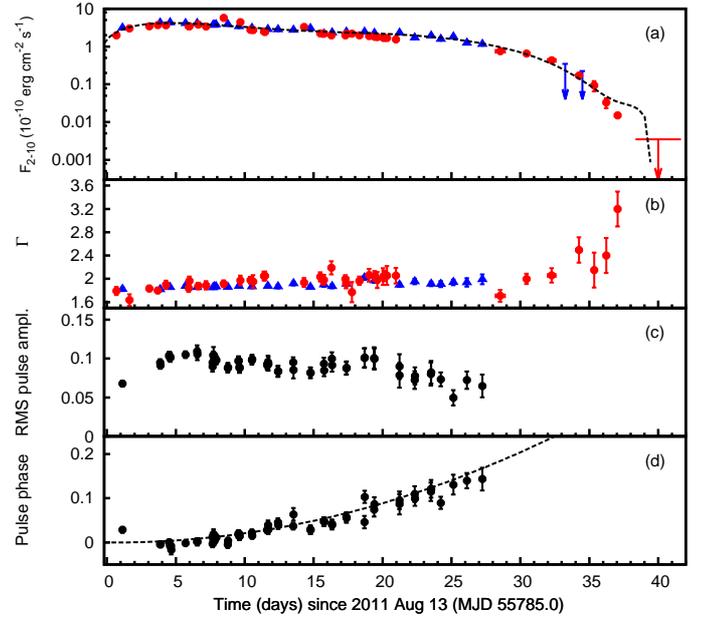}}
\caption{ Panel (a): absorbed 2--10 keV flux of {\fullsrc} as observed
  by the \textsl{Swift}/XRT (red circles) and by \textsl{RXTE}/PCA
  (blue triangles), and its best-fit sixth-order polynomial relation
  (dashed line).  The PCA fluxes were obtained after subtracting the
  spectrum observed during the interval 2011 Sep 20.5--22.4 (days 39.5
  and 41.4 in the scale used here), when the source already faded into
  quiescence; panel (b): photon index of the power-law spectra
  observed by \textsl{Swift}/XRT (red circles) and \textsl{RXTE}/PCA
  (blue triangles); panel (c): RMS amplitude of the 401 Hz signal as
  observed by the PCA; panel (d): pulse phases obtained by folding
  around $\nu_0=400.99018734$ Hz the PCA time-series corrected for the
  orbital motion by using the parameters listed in Table~\ref{tab},
  together with the best-fit quadratic model plotted as a dashed
  line.  }
\label{fig:lc}
\end{figure}

The 1--10 keV \textsl{Swift}/XRT spectra are satisfactorily modelled
by an absorbed power law. The best-fit value of the absorption column
ranges from 2.5 to 3.5$\times10^{22}$ cm$^{-2}$. Since these values
are compatible and much less accurate than the \textsl{Chandra}
estimate given by \citet{torres11},
n${\textrm{H}}=(2.87\pm0.04)\times10^{22}$ cm$^{-2}$, we fixed the
absorption column to this value in the fitting.  The 2--10 keV
absorbed flux is plotted as red circles in panel (a) of
Fig.~\ref{fig:lc}, together with a dashed line representing the
average flux trend obtained by fitting with a sixth-order polynomial
also the flux observed by the \textsl{RXTE}/PCA in the same energy
band (blue triangles; see below). We measured an excess of
$2.1(2)\times10^{-10}$ erg cm$^{-2}$ s$^{-1}$ with respect to this
trend during the \textsl{Swift} observation starting on 2011 Aug
20.5, a value significantly larger than the average
RMS scatter in the flux observed, $3.5\times10^{-11}$ erg cm$^{-2}$
s$^{-1}$.  The flux during this observation of
$F_{2-10}^{max}=(5.8\pm0.2)\times10^{-10}$ erg cm$^{-2}$ s$^{-1}$, is
the peak value observed during the outburst and corresponds to an
unabsorbed luminosity of $L_{2-10}^{peak}=(5.1\pm0.2)\times10^{36}$
d$_{7.6}^2$ erg s$^{-1}$, where $d_{7.6}$ is the distance to the
source in units of 7.6 kpc. An excess of $1.0(2)\times10^{-10}$ erg
cm$^{-2}$ s$^{-1}$ is also observed during the observation starting on
2011 Aug 21.5. By comparing these observations
with the nearest ones, a timescale as low as $\approx1000$ s is found
for the flux variability. Spectral variability is not observed as the
photon index, $\Gamma$, has values that are consistent with those
observed during most of the \textsl{Swift}/XRT observations (between
1.7 and 2; see red circles in panel (b) of Fig.~\ref{fig:lc}). A
steepening of the power law is instead observed as the transient fades
to quiescence. The source was no longer detected after Sep 19;
by co-adding the observations performed from Sep 19.4 to
22.7, we obtained an upper limit of $3.2\times10^{-13}$ erg
cm$^{-2}$ s$^{-1}$ to the 2--10 keV flux (3 $\sigma$ c.~l., assuming a
power-law with index $\Gamma=$1.8).

The X-ray transient {\fullsrc} lies very close to the central regions
of the Galactic bulge (l=0.1559$^{\circ}$, b=-1.0038$^{\circ}$) and
the emission of the Galactic ridge and numerous nearby sources
contaminate the observations in the field of view of the PCA, which is
about 1 by 1 degrees with a hexagonal shape. \citet{revnivtsev09}
estimated a 2--10 keV surface brightness of
$(8.6\pm0.5)\times10^{-11}$ erg cm$^{-2}$ s$^{-1}$ deg$^{-2}$ from a
nearby field not containing bright sources. Moreover, ten sources with
2--12 keV flux exceeding $5\times10^{-11}$ erg cm$^{-2}$ s$^{-1}$ were
found in a region of $1^{\circ}$ around the position of {\fullsrc} in
the fifth \textsl{XMM-Newton} serendipitous source catalogue
\citep{watson09}; the brightest of these sources is the persistent
burster 1A 1742--294 with a 2--10 keV flux between 2 and
4$\times10^{-10}$ erg cm$^{-2}$ s$^{-1}$. To estimate the
contamination of the PCA spectra, we considered observations performed
between 2011 Sep 20.5 and Sep 22.4, for an exposure of 3.7 ks, when
the transient had already faded into quiescence (see above). The net
count-rate observed by the top layer of the PCU2 during these
observations was 32.5(2) s$^{-1}$ (36.5(2) s$^{-1}$, if all the layers
are considered) and the absorbed 2--10 keV flux is
$(3.5\pm0.1)\times10^{-10}$ erg cm$^{-2}$ s$^{-1}$.  We considered the
spectrum obtained during these PCA observations as an additional
background to those performed previously.  After this subtraction, all
the PCA spectra could be accurately described by an absorbed
power-law, with the value of n$\textrm{H}$ fixed to the
\textsl{Chandra} estimate.  The 2--10 keV fluxes and the best-fit
values of the power-law photon index, ranging from 1.8 to 2, are
plotted as blue triangles in panel (a) and (b) of Fig.~\ref{fig:lc},
respectively. We found that the best-fit models of the simultaneous
\textsl{Swift}-XRT and \textsl{RXTE}-PCA spectra are consistent with each other.

The bolometric fluence of the outburst was estimated to be
$\mathcal{F}=(4.1\pm0.6)\times10^{-3}$ d$_{7.6}^2$ erg cm$^{-2}$ by
integrating the average trend of the 2--10 keV source flux (dashed
line in panel (a) of Fig.~\ref{fig:lc}) over the whole outburst
length, and scaling the value obtained by a \textsl{bolometric}
correction factor ($c_{bol}=6.0$) equal to the average observed ratio
of the unabsorbed flux extrapolated to the 0.05--100 keV band to the
absorbed flux observed in the 2--10 keV band. Assuming that $\Delta
M=R_* \mathcal{F}/GM_*$, the mass accreted by the NS during the
outburst is estimated to be $\Delta M=(8\pm1)\times10^{-11}$
d$_{7.6}^2$ M$_{\odot}$ for a $M_*=1.4$ M$_{\odot}$ NS with a radius
of $R_*=10$ km, a value typical of outbursts of AMSPs
\citep{galloway06}. However, these values can be overestimated by a
factor of $\sim2$ because of the unknown spectral distribution of the
source outside the energy band considered here. Describing the
broad-band spectrum with a Comptonization model with input photons and
hot scattering electrons with temperatures of $0.5$ and $50$ keV,
respectively, would rather yield a value of $c_{bol}\simeq3$.

\subsection{Timing analysis}
\label{sec:timing}

To perform a timing analysis of the 401 Hz signal, we reported the
photon arrival times to the barycentre of the Solar System by using
the \textsl{Chandra} position \citep{Chk11} and applied fine clock
corrections using the latest calibration file issued by the PCA team
at the time of the analysis (September 2011)\footnote{see
  http://heasarc.gsfc.nasa.gov/docs/xte/time\_news.html}, to reach an
accuracy in absolute timing of between 5 and 8 $\mu$s.

A signal at a frequency of between 400.95 and 401.0 Hz is observed in
power spectra produced over 256 s intervals, at a Lehay-normalised
power of $\approx 60$--$70$. A first solution of the orbital motion of
the NS was obtained by performing an epoch-folding search over 256 s
intervals and modelling the variations in the signal period caused by
the orbital motion. We thus obtained a first estimate of the orbital
parameters of the source, such as the orbital period,
$P_{orb}=$13835.65(5) s, the projected semi-major axis of the NS
orbit, $a\sin{i}/c=$0.3651(2) lt-s, the epoch of zero mean anomaly,
$T^*$, and upper limits to $e\sin{\omega}$ and $e\cos{\omega}$, where
$e$ is the eccentricity and $\omega$ is the longitude of the
periastron measured from the ascending node, as well as of the
barycentric spin frequency, $\nu_0=400.99020(2)$ Hz.

The times of arrival of X-ray photons were then corrected by using
this first orbital solution, and the time series folded into 16 phase
bins around the best estimate of the spin period, over 1500 s time
intervals. The presence and significance of pulsations in the folded
light curves were assessed following \citet{leahy83}.  The
background-subtracted 2--60 keV RMS amplitude of the observed profiles
is plotted in panel (c) of Fig.~\ref{fig:lc}. To take into account the
contribution of the the Galactic ridge and of nearby sources to the
PCA count rate, we considered an additional background level of
36.5(1) s$^{-1}$ (determined from PCU2 observations of the field, see
Sec.~\ref{sec:spectra}) for every active PCU. This is only an
approximation since the responses and background levels of the various
PCUs differ by about $10$ per cent, introducing an error in the
estimated amplitude that is comparable to the statistical
uncertainties in the various measurements. The intrinsic pulse RMS
amplitude thus evaluated takes values ranging from 0.06 to 0.11, with
relative uncertainties of the order of 10--15 per cent.  The pulse
profiles can generally be modelled by a sinusoid.  A second harmonic
with an RMS amplitude between 1 and 2 per cent, and a relative
uncertainty of $\approx 25$ per cent, is detected only during the
brightest observations.  Pulsations are not detected during the
observations performed on 2011 Sep 14 and 15 (MJD 55818--55819), with
an upper limit of about 20 per cent to the pulsed fraction, reflecting
our approximate knowledge of the true background level, especially
when the source flux is low.

\begin{table}
\caption{Spin and orbital parameters of {\fullsrc}.\label{tab}} \centering
\begin{tabular}{lr}
\hline
$\nu_0$ (Hz) & $400.99018734\pm1\times10^{-8}\pm8\times10^{-8}$   \\ 
$<\!\dot{\nu}\!>$  (Hz s$^{-1}$) & $(-6.3\pm1.1\pm1.5)\times10^{-14}$ \\
\hline
$a \sin{i}/c$ (lt-s) & 0.365165(5) \\
$P_{orb}$ (s) & 13835.619(1) \\
$T^*$ (MJD) & 55785.0600534(8) \\
$e$ & $<8\times10^{-5}$\\
$f(m_2,m_1,i)$ ($M_{\odot}$)& 0.00203807(8)\\
\hline
$\chi^2/\mbox{d.o.f.}$ & 76/50\\
\hline
\end{tabular}
\tablefoot{ The reference epoch for the solution is $\bar{T}=$ MJD
  55786.124. Numbers in parentheses are 1$\sigma$ errors in the last
  significant digit, while upper limits are evaluated at the 3$\sigma$
  confidence level. The uncertainties reported for $\nu_0$ and
  $<\!\dot{\nu}\!>$ are the statistical and the systematic uncertainty
  caused by the position error, respectively.  }
\end{table}

The estimates of the spin and orbital parameters of the pulsar were
refined  by applying  timing techniques to the pulse phase delays
\citep{deeter81}, i.e., modelling their evolution as
\begin{equation}
\label{eq:phases}
\phi(t)-\phi(\bar{T})=[\nu_0-\nu_F](t-\bar{T})+\frac{1}{2}<\!\dot{\nu}\!>(t-\bar{T})^2+R_{orb}(t),
\end{equation}
where $\bar{T}=$ MJD 55786.124 is the reference epoch of the folding,
$\nu_F$ is the folding frequency, $<\!\dot{\nu}\!>$ is the average
curvature of the trend followed by the phases, equal to the average
spin frequency derivative under the assumption that the spin is
well-tracked by the pulse phases, and $R_{orb}(t)$ describes a phase
modulation caused by the slight differences between the actual orbital
parameters of the system and those used to correct the time series,
i.e., $P_{orb}$, $a\sin{i}/c$, $T^*$, $e\sin\omega$, and
$e\cos{\omega}$. The best-fit parameters that we found are listed in
Table~\ref{tab}, while the pulse phases obtained by folding the {\it
  RXTE} light curves around $\nu_0=400.99018734$ Hz are plotted in
panel (d) of Fig.~\ref{fig:lc}, together with the best-fit model.  The
modeling of the pulse phase delays of {\fullsrc} with
Eq.~(\ref{eq:phases}) is satisfactory since $\chi^2/\mbox{d.o.f.}=
76/50\simeq1.5$.  We measured a statistically significant curvature of
the phase evolution $<\!\dot{\nu}\!>=(-6.3\pm1.1)\times10^{-14}$ Hz
s$^{-1}$, which is also indicated by the improvement in the
chi-squared of the model obtained by adding the quadratic term to a
linear model, $\Delta\chi^2=-47$. The interpretation of this curvature
in terms of a spin frequency derivative should indeed be taken with
caution. If the phases relative to the first \textsl{RXTE} observation
(spanning 2011 Aug 14.124--14.145; MJD 55786.124--55786.145) were
removed from the fit, the estimate of $<\!\dot{\nu}\!>$ would become
much less significant, $(-3.9\pm1.0)\times 10^{-14}$ Hz
s$^{-1}$. Moreover, the uncertainty in the source position introduces
a systematic error of the same order to this estimate.  By considering
the uncertainty in the position given by \citet{Chk11} thanks to an
observation perfomed by the HRC-S detector on-board \textsl{Chandra}
(0.6'' at 90\% confidence level, which corresponds to 0.4'' at 1
$\sigma$ c.l. according to the distribution of aspect offset of the
detector\footnote{see http://cxc.harvard.edu/cal/ASPECT/celmon/}), we
derived a systematic error in the estimates of the spin frequency and
of its derivative equal to $8\times10^{-8}$ Hz and $1.5\times10^{-14}$
Hz s$^{-1}$, respectively \citep{LynGsm90}.  Adding in quadrature this
error to the statistical uncertainty in $<\!\dot{\nu}\!>$, we found
that the detection of a frequency derivative is only marginally
significant. The modulus of the value obtained by fitting the whole
data-set, $<\!\dot{\nu}\!>=(-6.3\pm1.9)\times10^{-14}$ Hz s$^{-1}$, is
 3.3 times larger than its overall uncertainty, while the
estimate obtained by removing the first {\it RXTE} observation from
the sample becomes compatible with zero within the error.

\section{Discussion and conclusions}

We have presented an analysis of the observations performed by {\rxte} and
\textsl{Swift} during the outburst shown between August and September
2011 by the newly discovered AMSP {\fullsrc}. We have reported  the
discovery of a coherent signal at a frequency of 401 Hz in the {\rxte}
light curves, as well as derived an accurate orbital solution for the
3.84 hr binary system.

The pulse profile shown by {\fullsrc} is modelled by a sinusoid with
an RMS amplitude between 6 and 11 per cent.  A second harmonic with an
amplitude about ten times lower is detected only in a subset of
observations. Nearly sinusoidal profiles, with a similar ratio of the
second to the first harmonic amplitudes, had already been observed
from XTE J1751--305 \citep{Mrk02} and IGR J00291+5934
\citep{Gll05,Fln05}. The pulse phases shown by {\fullsrc} are quite
stable and their evolution can be closely described by a polynomial of
low ($\leq 2$) order, in addition to the delays introduced by the
orbital motion.  Interestingly, this property is shared by {\fullsrc}
with XTE J1751--305 \citep{Ppt08} and IGR J00291+5934
\citep{Fln05,Brd07,Ptr10,Hrt11,Ppt11d}, while the phases of other
AMSPs are instead affected by timing noise \citep[see,
  e.g.,][]{Hrt08,Rgg08}. We note how the phases of {\fullsrc} are
affected by an error of $\approx$0.01, a value consistent with the
phase uncertainty of a \textsl{noisy} pulsar such as SAX J1808.4-3658
(see Table 2 of \citealt{Hrt08}), such that the enhanced phase
stability observed in this case does not seem to be caused (at least
entirely) by a lower quality pulse statistics. The evolution of the
pulse phases of {\fullsrc} is best-fitted by a parabola with positive
curvature, indicating a NS spin-down at a rate of
$(-6.3\pm1.9)\times10^{-14}$ Hz s$^{-1}$ if interpreted in terms of
the NS spin frequency derivative. A similar trend was already observed
from other two AMSPs \citep{galloway02,papitto07}, but the low
significance of this measurement and the possible presence of residual
timing noise prevents us from drawing a firm detection of a NS spin
evolution.

\begin{figure}
\resizebox{\hsize}{!}{\includegraphics{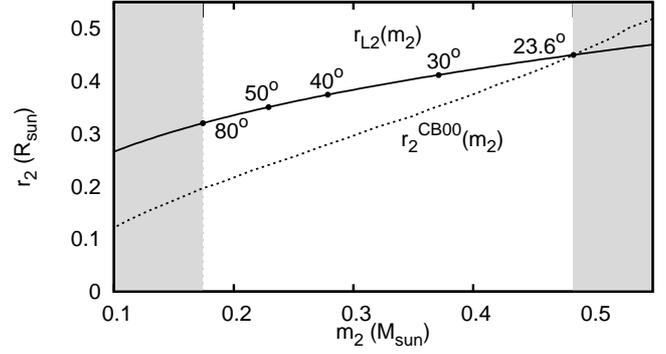}}
\caption{Size of the Roche lobe of the donor of {\fullsrc} (solid
  curve) and ZAMS mass-radius relation given by
  \citet[][dotted line]{ChbBrf00}. {The masses and radii of donor
    stars filling their Roche lobe, calculated from the measured mass
    function assuming $m_1=1.4$ and a set of values of inclination,
    are indicated as black points.} The left shaded region is excluded
  by the observed mass function and by the absence of eclipses (see
  text), while the right one is excluded by the condition that the
  companion star does not overfill its Roche lobe.}
\label{fig:orbit}
\end{figure}

The value of the pulsar mass function
$f(M_2,M_1,i)\simeq2\times10^{-3}$ M$_{\odot}$ allows us to set
constraints on the nature of the companion star. Since no eclipses are
observed during \textsl{RXTE}/PCA and \textsl{Swift}/XRT observations
performed when the NS is behind the companion (i.e., at true orbital
longitudes $\sim90^{\circ}$), we have inferred that the inclination of
the system is less than $80^{\circ}$. This provides a lower limit to
the companion mass of $m_2\simgt0.066+0.077 m_1$, where $m_1$ and
$m_2$ are the masses of the two stars in solar masses ($m_2\simgt0.17$
for $m_1=1.4$). An upper limit to the companion mass is obtained by
assuming that it does not overfill its Roche lobe, the size of which
is estimated by the relation of \citet{Pcz71}, $R_{L2}=0.462 a
[m_2/(m_1+m_2)]^{1/3}$, where $a$ is the binary separation. By using
the third Kepler law to relate the size of the orbit to both the total
mass of the system and the observed orbital period, one obtains $R_2
\simlt R_{L2}=0.573\: R_{\odot}\: m_2^{1/3}$. This relation is plotted
in Fig.~\ref{fig:orbit} together with the zero age main sequence
(ZAMS) mass-radius relation given by \citet{ChbBrf00}. By assuming
that the companion star fills its Roche Lobe and follows the ZAMS
mass-radius relation, we derived values of $m_2=0.48$ and $i=24.6^{\circ}$ for
the companion star mass and the system inclination,
respectively. While heavier companion stars can be excluded, lower
values are possible if the companion is bloated with respect to its
thermal equilibrium radius. {If a NS with a mass of 2 $M_{\odot}$ is
  considered, the lower limit to the system inclination set under
  these assumptions increases to 28.8$^{\circ}$. Among the scenarios
  explaining the evolution of LMXBs \citep[see][for a
    review]{deloye08}, the orbital period of {\fullsrc} and the limits
  to the mass of the companion derived here are consistent with a
  cataclysmic variable-like evolutionary path, with the low-mass donor
   that made contact with its Roche lobe before evolving  to
  the red giant branch \citep[see, e.g., the sequences plotted in
    Fig.~2 of][]{pods02}.}

\begin{acknowledgements}
 This work is supported by the Italian Space Agency, ASI-INAF
 I/088/06/0 and I/009/10/0 contracts for High Energy Astrophysics, and
 by the operating program of Regione Sardegna (ESF 2007-2013),
 L.R.7/2007, ``Promoting scientific research and technological
 innovation in Sardinia''. We thank the Swift team for the very rapid
 scheduling of the observations.
\end{acknowledgements}

\bibliographystyle{aa}
\bibliography{biblio}

\begin{thebibliography}{38}
\expandafter\ifx\csname natexlab\endcsname\relax\def\natexlab#1{#1}\fi

\bibitem[{{Altamirano} {et~al.}(2008){Altamirano}, {Casella}, {Patruno},
  {et~al.}}]{altamirano08}
{Altamirano}, D., {Casella}, P., {Patruno}, A., {et~al.} 2008, \apjl, 674, L45

\bibitem[{{Bozzo} {et~al.}(2011){Bozzo}, {Beardmore}, {Papitto},
  {et~al.}}]{Bzz11a}
{Bozzo}, E., {Beardmore}, A., {Papitto}, A., {et~al.} 2011, ATel, 3558, 1

\bibitem[{{Burderi} {et~al.}(2007){Burderi}, {Di Salvo}, {Lavagetto}, {Menna},
  {Papitto}, {Riggio}, {Iaria}, {D'Antona}, {Robba}, \& {Stella}}]{Brd07}
{Burderi}, L., {Di Salvo}, T., {Lavagetto}, G., {et~al.} 2007, \apj, 657, 961

\bibitem[{{Burrows} {et~al.}(2005){Burrows}, {Hill}, {Nousek},
  {et~al.}}]{burrows05}
{Burrows}, D.~N., {Hill}, J.~E., {Nousek}, J.~A., {et~al.} 2005, \ssr, 120, 165

\bibitem[{{Casella} {et~al.}(2008){Casella}, {Altamirano}, {Patruno},
  {et~al.}}]{casella08}
{Casella}, P., {Altamirano}, D., {Patruno}, A., {et~al.} 2008, \apjl, 674, L41

\bibitem[{{Chabrier} \& {Baraffe}(2000)}]{ChbBrf00}
{Chabrier}, G. \& {Baraffe}, I. 2000, \araa, 38, 337

\bibitem[{{Chakrabarty} {et~al.}(2011){Chakrabarty}, {Markwardt}, {Linares},
  {et~al.}}]{Chk11}
{Chakrabarty}, D., {Markwardt}, C.~B., {Linares}, M., {et~al.} 2011, ATel,
  3606, 1

\bibitem[{{Deeter} {et~al.}(1981){Deeter}, {Boynton}, \& {Pravdo}}]{deeter81}
{Deeter}, J.~E., {Boynton}, P.~E., \& {Pravdo}, S.~H. 1981, \apj, 247, 1003

\bibitem[{{Deloye}(2008)}]{deloye08}
{Deloye}, C.~J. 2008, in American Institute of Physics Conference Series, Vol.
  983, 40 Years of Pulsars: Millisecond Pulsars, Magnetars and More, ed.
  {C.~Bassa, Z.~Wang, A.~Cumming, \& V.~M.~Kaspi}, 501--509

\bibitem[{{Falanga} {et~al.}(2005){Falanga}, {Kuiper}, {Poutanen}, {Bonning},
  {Hermsen}, {di Salvo}, {Goldoni}, {Goldwurm}, {Shaw}, \& {Stella}}]{Fln05}
{Falanga}, M., {Kuiper}, L., {Poutanen}, J., {et~al.} 2005, \aap, 444, 15

\bibitem[{{Ferrigno} {et~al.}(2011){Ferrigno}, {Bozzo}, {Gibaud},
  {et~al.}}]{Frr11}
{Ferrigno}, C., {Bozzo}, E., {Gibaud}, L., {et~al.} 2011, ATel, 3560, 1

\bibitem[{{Galloway}(2006)}]{galloway06}
{Galloway}, D.~K. 2006, in American Institute of Physics Conference Series,
  Vol. 840, The Transient Milky Way: A Perspective for MIRAX, ed. {F.~D'Amico,
  J.~Braga, \& R.~E.~Rothschild}, 50--54

\bibitem[{{Galloway} {et~al.}(2002){Galloway}, {Chakrabarty}, {Morgan},
  {et~al.}}]{galloway02}
{Galloway}, D.~K., {Chakrabarty}, D., {Morgan}, E.~H., {et~al.} 2002, \apjl,
  576, L137

\bibitem[{{Galloway} {et~al.}(2005){Galloway}, {Markwardt}, {Morgan},
  {et~al.}}]{Gll05}
{Galloway}, D.~K., {Markwardt}, C.~B., {Morgan}, E.~H., {et~al.} 2005, \apjl,
  622, L45

\bibitem[{{Gehrels} {et~al.}(2004){Gehrels}, {Chincarini}, {Giommi},
  {et~al.}}]{gehrels04}
{Gehrels}, N., {Chincarini}, G., {Giommi}, P., {et~al.} 2004, \apj, 611, 1005

\bibitem[{{Gibaud} {et~al.}(2011){Gibaud}, {Bazzano}, {Bozzo},
  {et~al.}}]{Gbd11}
{Gibaud}, L., {Bazzano}, A., {Bozzo}, E., {et~al.} 2011, ATel, 3551, 1

\bibitem[{{Hartman} {et~al.}(2011){Hartman}, {Galloway}, \&
  {Chakrabarty}}]{Hrt11}
{Hartman}, J.~M., {Galloway}, D.~K., \& {Chakrabarty}, D. 2011, \apj, 726, 26

\bibitem[{{Hartman} {et~al.}(2008){Hartman}, {Patruno}, {Chakrabarty},
  {Kaplan}, {Markwardt}, {Morgan}, {Ray}, {van der Klis}, \&
  {Wijnands}}]{Hrt08}
{Hartman}, J.~M., {Patruno}, A., {Chakrabarty}, D., {et~al.} 2008, \apj, 675,
  1468

\bibitem[{{Jahoda} {et~al.}(2006){Jahoda}, {Markwardt}, {Radeva},
  {et~al.}}]{jahoda06}
{Jahoda}, K., {Markwardt}, C.~B., {Radeva}, Y., {et~al.} 2006, \apjs, 163, 401

\bibitem[{{Leahy} {et~al.}(1983){Leahy}, {Darbro}, {Elsner}, {Weisskopf},
  {Kahn}, {Sutherland}, \& {Grindlay}}]{leahy83}
{Leahy}, D.~A., {Darbro}, W., {Elsner}, R.~F., {et~al.} 1983, \apj, 266, 160

\bibitem[{{Linares} {et~al.}(2011{\natexlab{a}}){Linares}, {Altamirano},
  {Watts}, {et~al.}}]{Lnr11}
{Linares}, M., {Altamirano}, D., {Watts}, A., {et~al.} 2011{\natexlab{a}},
  ATel, 3568, 1

\bibitem[{{Linares} {et~al.}(2011{\natexlab{b}}){Linares}, {Bozzo},
  {Altamirano}, {et~al.}}]{linares11}
{Linares}, M., {Bozzo}, E., {Altamirano}, D., {et~al.} 2011{\natexlab{b}},
  ATel, 3661, 1

\bibitem[{{Lyne} \& {Graham-Smith}(1990)}]{LynGsm90}
{Lyne}, A.~G. \& {Graham-Smith}, F. 1990, {Pulsar astronomy (Cambridge
  University Press, Cambridge)}

\bibitem[{{Markwardt} {et~al.}(2011){Markwardt}, {Strohmayer}, \&
  {Smith}}]{Mrk11}
{Markwardt}, C.~B., {Strohmayer}, T.~E., \& {Smith}, E.~A. 2011, ATel, 3561, 1

\bibitem[{{Markwardt} {et~al.}(2002){Markwardt}, {Swank}, {Strohmayer},
  {et~al.}}]{Mrk02}
{Markwardt}, C.~B., {Swank}, J.~H., {Strohmayer}, T.~E., {et~al.} 2002, \apjl,
  575, L21

\bibitem[{{Paczy{\'n}ski}(1971)}]{Pcz71}
{Paczy{\'n}ski}, B. 1971, \araa, 9, 183

\bibitem[{{Papitto} {et~al.}(2011{\natexlab{a}}){Papitto}, {Belloni},
  {Ferrigno}, {et~al.}}]{Ppt11b}
{Papitto}, A., {Belloni}, T.~M., {Ferrigno}, C., {et~al.} 2011{\natexlab{a}},
  ATel, 3563, 1

\bibitem[{{Papitto} {et~al.}(2007){Papitto}, {Di Salvo}, {Burderi}, {Menna},
  {Lavagetto}, \& {Riggio}}]{papitto07}
{Papitto}, A., {Di Salvo}, T., {Burderi}, L., {et~al.} 2007, \mnras, 375, 971

\bibitem[{{Papitto} {et~al.}(2011{\natexlab{b}}){Papitto}, {Ferrigno}, {Bozzo},
  {et~al.}}]{Ppt11a}
{Papitto}, A., {Ferrigno}, C., {Bozzo}, E., {et~al.} 2011{\natexlab{b}}, ATel,
  3556, 1

\bibitem[{{Papitto} {et~al.}(2008){Papitto}, {Menna}, {Burderi},
  {et~al.}}]{Ppt08}
{Papitto}, A., {Menna}, M.~T., {Burderi}, L., {et~al.} 2008, \mnras, 383, 411

\bibitem[{{Papitto} {et~al.}(2011{\natexlab{c}}){Papitto}, {Riggio}, {Burderi},
  {di Salvo}, {D'A{\'{\i}}}, \& {Iaria}}]{Ppt11d}
{Papitto}, A., {Riggio}, A., {Burderi}, L., {et~al.} 2011{\natexlab{c}}, \aap,
  528, A55+

\bibitem[{{Patruno}(2010)}]{Ptr10}
{Patruno}, A. 2010, \apj, 722, 909

\bibitem[{{Podsiadlowski} {et~al.}(2002){Podsiadlowski}, {Rappaport}, \&
  {Pfahl}}]{pods02}
{Podsiadlowski}, P., {Rappaport}, S., \& {Pfahl}, E.~D. 2002, \apj, 565, 1107

\bibitem[{{Revnivtsev} {et~al.}(2009){Revnivtsev}, {Sazonov}, {Churazov},
  {Forman}, {Vikhlinin}, \& {Sunyaev}}]{revnivtsev09}
{Revnivtsev}, M., {Sazonov}, S., {Churazov}, E., {et~al.} 2009, \nat, 458, 1142

\bibitem[{{Riggio} {et~al.}(2008){Riggio}, {Di Salvo}, {Burderi}, {Menna},
  {Papitto}, {Iaria}, \& {Lavagetto}}]{Rgg08}
{Riggio}, A., {Di Salvo}, T., {Burderi}, L., {et~al.} 2008, \apj, 678, 1273

\bibitem[{{Torres} {et~al.}(2011){Torres}, {Madej}, {Jonker},
  {et~al.}}]{torres11}
{Torres}, M.~A.~P., {Madej}, O., {Jonker}, P.~G., {et~al.} 2011, ATel, 3638, 1

\bibitem[{{Watson} {et~al.}(2009){Watson}, {Schr{\"o}der}, {Fyfe},
  {et~al.}}]{watson09}
{Watson}, M.~G., {Schr{\"o}der}, A.~C., {Fyfe}, D., {et~al.} 2009, \aap, 493,
  339

\bibitem[{{Wijnands} \& {van der Klis}(1998)}]{WijvdK98}
{Wijnands}, R. \& {van der Klis}, M. 1998, \nat, 394, 344

\end{thebibliography}

\end{document}